\documentclass[prl,twocolumn]{revtex4}

\usepackage{amssymb,amsthm}
\usepackage{mathtools}

\newcommand{\cM}{{\cal M}}
\newcommand{\id}{{\mathbb I}}
\newcommand{\im}{{\rm i }}
\newcommand{\m}{{\bf m }}

\newcommand\be{\begin{eqnarray}}
\newcommand\ee{\end{eqnarray}}

\begin{document}

\title{General Relativity from Three-Forms in Seven Dimensions}
\author{Kirill Krasnov} 
\affiliation{School of Mathematical Sciences, University of Nottingham, UK}

\begin{abstract} We consider a certain theory of 3-forms in 7 dimensions, and study its dimensional reduction to 4D, compactifying the 7-dimensional manifold on the 3-sphere of a fixed radius. We show that the resulting 4D theory is General Relativity (GR) in Plebanski formulation, modulo corrections that are negligible for curvatures smaller than Planckian. Possibly the most interesting point of this construction is that the dimensionally reduced theory is GR with a non-zero cosmological constant, and the value of the cosmological constant is directly related to the size of $S^3$. Realistic values of $\Lambda$ correspond to $S^3$ of Planck size. 
\end{abstract}

\date{November 2016}
\maketitle

The fundamental fact about a generic 3-form $C$ on a 7-dimensional manifold $\cM$ is that it defines a metric $g_C$. The metric is explicitly given by the following formula
\be\label{metric}
g_C(\xi,\eta) {\rm vol}_C = -\frac{1}{6} i_\xi C\wedge i_\eta C\wedge C.
\ee
Here $g_C(\xi,\eta)$ is the result of the metric contraction of two vector fields $\xi,\eta$, ${\rm vol}_C$ is the volume form for $g_C$, and $i_{\xi,\eta}$ is the operation of insertion of a vector field into a form. The minus sign in this formula is convention dependent, see below for ours. The metric (\ref{metric}) has been known for more than a century, see e.g. \cite{History} for a historical perspective. It is ultimately related to the geometry of spinors in 7 and 8 dimensions, see e.g. \cite{Friedrich:1995dp} for the discussion of the spinor aspect, and to octonions, see e.g. \cite{Walpuski}. 

Generic 3-forms in 7 dimensions are related to the exceptional group $G_2$. This can be defined as the subgroup of ${\rm GL}(7)$ that stabilizes a generic 3-form, see \cite{Bryant} and more recently \cite{Hitchin:2001rw}. The space of generic 3-forms (at a point) can then be identified with the coset ${\rm GL}(7)/G_2$. The fact that $C$ defines $g_C$ explains why $G_2$ is a subgroup of ${\rm SO}(7)$. 

The volume form ${\rm vol}_C$, playing an important role below, can also be described explicitly as a homogeneity degree $7/3$ object built from $C$. Thus, let $\tilde{\epsilon}^{a_1\dots a_7}$ be the densitiesed completely anti-symmetric tensor taking values $\pm 1$ in any coordinate system. Here $a=1,\ldots,7$. We can then construct the following degree 7 and weight 3 scalar:
\be\label{volume}
\tilde{\epsilon}^{a_1\dots a_7}\tilde{\epsilon}^{b_1\dots b_7}\tilde{\epsilon}^{c_1\dots c_7}
C_{a_1b_1c_1} \ldots C_{a_7 b_7 c_7}.
\ee
The cube root of this expression is a multiple of ${\rm vol}_C$. This is not the most useful in practice way of computing the volume form -- it is usually much more effective to compute the volume ${\rm vol}_C$ from the determinant of $g_C$. But it is comforting to know that the explicit expression (\ref{volume}) is possible. 

Let us now make $C$ dynamical. Consider the following action principle
\be\label{action}
S[C] = \frac{1}{2} \int_\cM C\wedge dC + 6\lambda \,{\rm vol}_C.
\ee
The first term here (i.e. the case $\lambda=0$) describes a topological field theory considered in \cite{Gerasimov:2004yx}. The Euler-Lagrange equations following from (\ref{action}) are
\be\label{feqs}
dC = \lambda \, {}^*C.
\ee
Here ${}^*C$ is the Hodge dual of $C$ computed using $g_C$. The numerical coefficient on the right-hand-side here is simplest verified by noticing that ${\rm vol}_C = -(1/7) C\wedge {}^*C$, and then using the homogeneity to compute the variation of ${\rm vol}_C$ with respect to $C$. We note that, because the two terms in (\ref{action}) scale differently, by rescaling $C$ we can always achieve $\lambda=1$ at the expense of introducing a parameter in front of the action. We will do so from now on. Thus, there are no free parameters in the theory (\ref{action}). 

Real 3-forms $C$ are of two possible types. Forms of one type give $g_C$ of signature $(4,3)$. Forms of the other type give the Riemannian signature metrics $g_C$. Such 3-forms satisfying (\ref{feqs}) describe what \cite{Friedrich:1995dp} call nearly parallel $G_2$ structures. Note that (\ref{feqs}) implies that ${}^*C$ is closed. However, this equation also says that $dC\not=0$. Thus, the critical points of (\ref{action}) are {\it not} the possibly more familiar in this context torsion-free $G_2$ structures satisfying $dC=0, d{}^*C=0$ and describing $G_2$ holonomy manifolds. A related observation is that the equation (\ref{feqs}) implies that the metric $g_C$ is Einstein with non-zero scalar curvature, see proposition 3.10 from \cite{Friedrich:1995dp}. In contrast, $G_2$-holonomy manifolds are Ricci flat, see e.g. \cite{Hitchin:2001rw}. 

While equations (\ref{feqs}) have been studied in the literature, the variational principle (\ref{action}) is new. Note that action (\ref{action}) is different from the ones considered by Hitchin \cite{Hitchin:2001rw}. The simplest Hitchin action is the last term in (\ref{action}), restricted to 3-forms in a fixed cohomology class. Our action is the sum of those in \cite{Gerasimov:2004yx} and \cite{Hitchin:2001rw}, with no constraint on $C$. 

We now describe a relation to 4D General Relativity (GR). We claim that (\ref{action}) dimensionally reduced on $S^3$ (of a fixed radius) is a 4D theory of gravity that is for all practical purposes indistinguishable from GR. This means that while the reduced theory is, strictly speaking, not GR, it coincides with GR for Weyl curvatures smaller than Planckian, which is anyway the regime where we can trust GR as a classical theory. All this is to be explained in more details below.  

To explain why the outlined embedding of 4D GR into a theory of 3-forms in 7D may be interesting, let us remind the reader the basics of Kaluza-Klein (KK) theory. Here one starts with GR in higher dimensions and dimensionally reduces to 4 dimensions. In the simplest and also historically the first setup one starts with GR (with zero cosmological constant) in 5 dimensions and dimensionally reduces on $S^1$. If one fixes the size of $S^1$, as was done in the first treatments, the dimensionally reduced 4D theory is GR coupled to Maxwell. Allowing the size of the circle to become dynamical gives rise to an additional massless scalar field in 4D, and to avoid conflict with observations this must be given a mass, or stabilised in some other way. 

The Kaluza-Klein mechanism gives a geometrically compelling unification of gravity with electromagnetism. Also, as pointed out by Kaluza, it relates the quantum of electric charge to the size of the compact extra dimension. It can be generalised to non-Abelian gauge fields. A comprehensive review on KK is e.g. \cite{Duff:1986hr}.

Let us return to our story. We claimed that 4D GR arises as the dimensional reduction of the theory (\ref{action}) of 3-forms in 7D. Unlike the KK case, no unification is achieved here, the reduced theory is pure gravity. Also unlike KK, gravity in 4D arises from a 7D theory of a very different sort -- the theory (\ref{action}) is a dynamical theory of 3-forms, not metrics. While it may be amusing that 4D GR admits a lift to a theory of such a different nature as (\ref{action}), is this a useful perspective on 4D gravity? 

Now comes what we believe is the key point of our construction. As we will show, the dimensionally reduced theory is GR with non-zero cosmological constant, and the value of the cosmological constant is directly related to the size of the $S^3$. As the 5D Kaluza-Klein story makes the electric charge a dynamically determinably quantity, at least in principle, via some "spontaneous compactification" mechanism, in our setup the 4D cosmological constant becomes in principle determinable by the dynamics of the extra dimensions. 

Thus, our 7D lift of 4D GR makes the 4D cosmological constant a dynamical object, at least in principle determinable by the dynamics of the extra dimensions. Having said this, we must also say that in this short paper we limit ourselves to just demonstrating the relation between the radius of $S^3$ and $\Lambda$. No attempt at studying the dynamics of the extra dimensions (and thus predicting $\Lambda$) will be made. Still, this should be kept in mind as the strongest motivation for our construction. 

After these motivational remarks, we are ready to describe the dimensional reduction. We phrase the discussion that follows in terms of real objects. In this case the dimensionally reduced theory is the Riemannian signature GR. All objects can also be complexified, in this case one obtains complexified GR. The subtler issue of reality conditions relevant for the Lorentzian signature theory will be described elsewhere. 

We assume that the group ${\rm SU}(2)$ acts on $\cM$ freely. This gives $\cM$ the structure of an ${\rm SU}(2)$ principal bundle over a 4-dimensional base $M$. Our considerations here are local, over a region in $M$. We parametrise fiber points as $g\in{\rm SU}(2)$, with the group action on the fiber being the right action of ${\rm SU}(2)$ on itself. Denote by $\m=g^{-1} dg$ the Maurer-Cartan one-forms on ${\rm SU}(2)$. To establish notations, let $\bf A$ be an ${\rm SU}(2)$ connection on the base $M$, i.e. a $2\times 2$ anti-Hermitian matrix valued one-form on $M$, and let $A=g^{-1} {\bf A} g$ be its lift into the total space of the bundle. Then $W=\m+A$ is the connection one-form in the total space of the bundle. Simple standard computation shows that $F:=dW+WW$ is a 2-form that is purely horizontal $F=g^{-1} {\bf F} g$, where ${\bf F}=d{\bf A}+{\bf A}{\bf A}$ is the curvature of the connection one-form on the base. Here and in what follows, for brevity, we omit the wedge product symbol. 

A general ${\rm SU}(2)$ invariant 3-form on $X$ can be written as $C= {\rm Tr}(\phi \m^3 + A \m^2 + B \m) + c$. Here $\phi\in \Lambda^0(M), c\in\Lambda^3(M)$, while $A,B$ are lifts to the bundle of Lie algebra valued 1- and 2-forms on the base $M$ respectively. Note that none of the 35 components of $C$ has been lost here, as a simple count of components in $\phi,A,B,c$ shows. The above parametrisation of $C$ is however not the one most suited for computations. We note that the terms qubic and quadratic in $\m$ can always be combined, at the expense of redefining the other fields. This suggest we parametrise 
\be\label{C*}
C = - 2 \,{\rm Tr} \left( \frac{1}{3} \phi^3 W^3 + \phi W B\right) + c.
\ee
Here $W=\m+A$ is a connection in the total space of the bundle. The objects $\phi,A,B,c$ appearing here are different from those above, but of the same nature. The parametrisation (\ref{C*}) is most suited for practical computations. Numerical prefactors are for future convenience. 

A simple computation then gives
\be
dC= - 2 \,{\rm Tr} \Big( \phi^2 d\phi W^3 +( \phi^3 F+ \phi B) W^2 \\ \nonumber 
+( d\phi B + \phi d_A B)W + \phi FB \big) + dc.
\ee
Here $d_A B = g^{-1} (d{\bf B}+{\bf A}{\bf B}-{\bf B}{\bf A}) g$ is the lift to the bundle of the covariant derivative of Lie algebra-valued 2-form ${\bf B}$ with respect to the connection $\bf A$. Another simple computation using some trace identities gives
\be\label{cdc}
\frac{1}{2} \int_\cM C dC = \int_{{\rm SU}(2)} -\frac{2}{3} {\rm Tr}(\m^3) \\ \nonumber \times
\int_M -2\, {\rm Tr}(\phi^4 {\bf B}{\bf F} + (\phi^2/2) {\bf B}{\bf B}) + \phi^3 dc.
\ee
We learn that the dimensional reduction of the first, topological term in (\ref{action}), modulo the prefactor equal to the volume of ${\rm SU}(2)$, is the so-called BF theory with a $\Lambda$-term, coupled to the scalar and 3-form fields. We find this result interesting in its own right. The dimensional reduction of the topological theory is topological. Thus, if there is no second term in (\ref{action}), varying with respect to $c$ gives $\phi=const$, and we recover the usual Lagrangian of the topological BF theory with the $\Lambda$ term. 

Let us now understand the dimensional reduction of the second term in (\ref{action}). This is a no-derivative term, so it only changes the "potential" for the $\phi,B,c$ fields. In this paper we will set $\phi=const$ and $c=0$. The complete dimensional reduction will be carried out elsewhere. Setting the size of the extra dimensions to a constant is achieved by $\phi=const$. At the same time, it is clear from (\ref{cdc}) that the 3-form field $c$ is "conjugate" to $\phi$ and so setting this field to constant justifies setting $c$ to zero. 

To compute the volume form corresponding to (\ref{C*}) (with $c=0$) we need to write this 3-form in ${\rm SO}(3)$ notations. This is achieved by decomposing all matrix-valued fields in terms of the ${\rm SU}(2)$ generators $\tau^i = -(\im/2) \sigma^i$, where $\sigma^i$ are the usual Pauli matrices. So, we write $W=W^i\tau^i$ etc. This gives
\be\label{c-so3}
C= \frac{\phi^3}{6} \epsilon^{ijk} W^i W^j W^k + \phi W^i B^i.
\ee
The metric $g_C$ and thus the volume form ${\rm vol}_C$ are then easiest computed by putting this $C$ into its canonical form. As the canonical form we take
\be\label{c-canon}
C= \frac{1}{6} \epsilon^{ijk} e^i e^j e^k + e^i \Sigma^i.
\ee
Here $\Sigma^i$ is the basis of anti-self-dual 2-forms
\be\nonumber
\Sigma^1= e^{45}-e^{67}, \quad \Sigma^2 = e^{46}-e^{75}, \quad \Sigma^3=e^{47}-e^{56}.
\ee
The notation here is $e^{ab}=e^a e^b$. It is then easy to check that for $C$ in its canonical form (\ref{c-canon}), the metric defined by $C$ via (\ref{metric}) is $ds^2_C= \sum_{a=1}^7 (e^a)^2$. 

To compute the metric for (\ref{c-so3}) we need to rewrite it in the canonical form (\ref{c-canon}). This is done by choosing a convenient parametrisation of $B^i$ fields. To establish this parametrisation, we note that the triple of 2-forms $B^i$ defines a metric on the base in which these forms are anti-self-dual (ASD). This is the Urbantke metric \cite{Urbantke}. In fact, a simple calculation with the formula (\ref{metric}) shows that the Urbantke formula 
\be
g_\Sigma(\xi,\eta) {\rm vol}_\Sigma = -\frac{1}{6} \epsilon^{ijk} i_\xi \Sigma^i \wedge i_\eta \Sigma^j \wedge \Sigma^k
\ee
arises as the metric on the base from (\ref{metric}), with $C$ in its canonical form (\ref{c-canon}). This clearly points towards a 7-dimensional origin of the Urbantke formula. The 2-forms $B^i$ can then always be parametrised as 
\be\label{B-X}
B^i = \sqrt{X}^{ij} \Sigma^j.
\ee
Here $\Sigma^i$ is an orthonormal basis of ASD 2-forms for the metric defined (via Urbantke formula) by $B^i$. The matrix $X^{ij}$ is defined as that of the wedge products of $B^i$. We have $B^i\wedge B^j = - 2 X^{ij} {\rm vol}_\Sigma$, where ${\rm vol}_\Sigma$ is the volume form of the metric whose ASD 2-forms are $\Sigma^i$. Substituting the parametrisation (\ref{B-X}) into (\ref{c-so3}) we see that the 3-form can be written in the following way
\be
C= \rho \left( \frac{1}{6} \epsilon^{ijk} e^i e^j e^k + e^i \Sigma^i\right),
\ee
with
\be
\rho = \left({\rm det}(X)\right)^{1/4}, \quad e^i = \frac{\phi}{\rho} \sqrt{X}^{ij} W^j.
\ee
This puts $C$ into a form that is a multiple of the canonical. The metric $g_C$ is then $\rho^{2/3}$ time the metric for which the above $e^a$ is the frame. This gives
\be
ds^2_C = \phi^2 W^i \frac{X^{ij}}{({\rm det}(X))^{1/3}} W^j + ({\rm det}(X))^{1/6} ds^2_\Sigma.
\ee
The volume form for this metric is
\be
{\rm vol}_C= \frac{\phi^3}{6}\epsilon^{ijk} W^i W^j W^k ({\rm det}(X))^{1/3} {\rm vol}_\Sigma.
\ee

We now put all pieces together and write the dimensionally reduced 4D Lagrangian, which is (\ref{action}) on the ansatz (\ref{C*}) (with $c=0$), divided by the volume of the fiber. We have
\be\label{eff-L}
L_{{\rm 4D}} = \phi^4 B^i F^i + \frac{\phi^2}{2} B^i B^i + 3\phi^3 ({\rm det}(X))^{1/3} {\rm vol}_\Sigma.
\ee
This is a Lagrangian of the type "BF theory plus a potential for the B field". From general considerations in \cite{Krasnov:2008fm} we known that this is a 4D gravity theory. 

We would now like to show how this theory reduces to GR. To this end, let us rewrite the last term in (\ref{eff-L}) by introducing an auxiliary matrix field. We have
\be\nonumber
6({\rm det}(X))^{1/3} {\rm vol}_\Sigma = - H^{ij} B^i B^j  +2\mu ( {\rm det}(H) - 1){\rm vol}_\Sigma.
\ee
Indeed, varying the right-hand-side with respect to $H^{ij}$ we get $X^{ij} = \mu\, {\rm det}(H) (H^{-1})^{ij}$. The condition ${\rm det}(H)=1$ imposed by the Lagrange multiplier $\mu$ then sets $\mu =( {\rm det}(X))^{1/3}$. Substituting the resulting solution $H^{ij} = ( {\rm det}(X))^{1/3} (X^{-1})^{ij}$ into the first term we reproduce the left-hand-side. 

Using the above way of writing the last term in (\ref{eff-L}) we can rewrite the 4D Lagrangian as follows
\be\label{L*}
L_{{\rm 4D}} /\phi^4 = B^i F^i - \frac{1}{2} M^{ij} B^i B^j \\ \nonumber
+2\mu ({\rm det}(\id + \phi^2 M) - \phi^3){\rm vol}_\Sigma.
\ee
Here we defined a new matrix $M^{ij}$ so that $\phi H= \id + \phi^2 M$, and redefined the Lagrange multiplier $\mu$. The key point now is that the constraint ${\rm det}(\id + \phi^2 M)=const$, when expanded in powers of $M$, is the constraint ${\rm Tr}(M)=const$, and this is known to give General Relativity in its Plebanski formulation \cite{Plebanski:1977zz}, \cite{Krasnov:2009pu}. 

We thus claim that (\ref{L*}) describes GR, plus higher order corrections immaterial in the regime of not too high Weyl curvatures. Let us make all this more precise. To this end, we reparametrise (\ref{L*}) by writing 
\be\label{M}
M^{ij} = \Psi^{ij}+ \frac{1}{3} \Lambda(\Psi).
\ee
Here $\Psi^{ij}$ is the tracefree part of $M^{ij}$ and $\Lambda(\Psi)$ is the function to be found by imposing the constraint in the second line in (\ref{L*}). This is the parametrisation in which this class of 4D gravity theories was discovered in \cite{Krasnov:2006du}. The constraint reads
\be
\left( 1+ \frac{\Lambda\phi^2}{3}\right)^3 - \frac{3}{2} \left( 1+ \frac{\Lambda\phi^2}{3}\right) \phi^4 {\rm Tr}(\Psi^2) \\ \nonumber + \phi^6 {\rm det}(\Psi) = \phi^3.
\ee
We can now solve for $\Lambda=\Lambda(\Psi)$ as a power series expansion, under assumption that $\Psi\ll 1$. To order $\Psi^2$ 
\be\label{Lambda}
\frac{\Lambda(\Psi)}{3} = \frac{\phi-1}{\phi^2} + \frac{\phi}{2} {\rm Tr}(\Psi^2) + O(\Psi^3). 
\ee
To see that this is indistinguishable from GR, we remind the reader the Plebanski Lagrangian \cite{Plebanski:1977zz}, see also \cite{Krasnov:2009pu}
\be
L'_{\rm Pleb} = M_p^2 \left( B^i F^i - \frac{1}{2} \left(\Psi^{ij} + \frac{\Lambda}{3}\right) B^i B^j\right).
\ee
Here $M_p^2 = 1/8\pi G$ is the Planck mass, $G$ is the Newton's, and $\Lambda$ is the cosmological constant. Here $B^i$ is a dimensionless field that describes the metric (via Urbantke formula). We now absorb the Planck mass so as to make $B^i$ (and thus the metric) dimensionful. The dimensionful metric measures distances in units of the Planck length. Thus, we redefine $B^i \to B^i/M_p^2, \Psi\to M_p^2 \Psi, \Lambda\to M_p^2 \Lambda$. 
\be\label{Pleb}
L_{\rm Pleb} = B^i F^i - \frac{1}{2} \left(\Psi^{ij} + \frac{\Lambda}{3}\right) B^i B^j.
\ee
The new $\Psi, \Lambda$ are dimensionless. The object $\Psi$ is the (anti-self-dual part of) the Weyl curvature, measured in Planck units. So, it satisfies $\Psi\ll 1$ in all situations in which GR has been tested, or can be trusted. 

Coming back to (\ref{L*}), for $\Psi\ll 1$ the dependence of $\Lambda(\Psi)$ on $\Psi$ in (\ref{Lambda}) can be neglected and $\Lambda(\Psi)$ becomes a constant. This shows that the theory (\ref{L*}) is indistinguishable from GR in its form (\ref{Pleb}) in all situations where GR has been tested and/or can be trusted. Note that the $O(\Psi^2)$ term in (\ref{Lambda}) is neglected as compared to $\Psi^{ij}$ term in (\ref{M}), not as compared to the constant, which can be small. Detailed study of effects of modification such as (\ref{Lambda}) on the spherically symmetric solution of GR can be found in \cite{Krasnov:2007ky}.

The dimensionless $\Lambda$ is the cosmological constant measure in Planck units, and is the extraordinary small number $\Lambda\sim 10^{-120}$ that embodies the cosmological constant problem. Our theory (\ref{L*}) gives small cosmological constant for values of radius of compactification $\phi$ close to unity. This must hold to extraordinary high accuracy
\be\label{phi-Lambda}
\phi -1= \frac{\Lambda}{3M_p^2},
\ee
where we now reinstated the Planck mass so that this is the usual dimensionful $\Lambda$, and omitted higher order terms. In (\ref{Lambda}) one can also get small $\Lambda$ for large $\phi$, but presumably this should be deemed unphysical. 

To summarise, we have shown that the dimensionally reduced theory (\ref{action}) gives a 4D gravity theory that in the regime of small (as compared to Planckian) Weyl curvatures $\Psi\ll M_p^2$ is indistinguishable from GR in Plebanski formulation. This can be stated as
\be
L_{\rm 4D}\approx L_{\rm Pleb},
\ee
with the relation between the radius $\phi$ of $S^3$ and $\Lambda$ given by (\ref{phi-Lambda}). The relation (\ref{phi-Lambda}),  together with the dimensionless value $\Lambda$ being so small predicts that the fibers $S^3$ are of Planck size. This is what is usually expected from compactified extra dimensions, which we find gratifying. 

There are many things that need to be done to convert the model studied here into a realistic theory of gravity. First and foremost, one must see whether there is a dynamical mechanism for driving the compactification radius to unity. As we have seen, such a mechanism is necessary to explain the smallness of $\Lambda$ in our approach. Second, one must study how to describe Lorentzian signature gravity in this framework. It may well be that the two questions are not unrelated. It is clear that the size of $S^3$, from the point of view of 4D, behaves as a scalar field. It is interesting to study the dynamics of this scalar field in the "Early Universe", with the question being whether this field can play the role of the inflaton. Finally, there is also the question of coupling to matter, but this can probably be postponed till one unravels all the consequences of this model of "empty" Universe. 

Finally, to avoid confusion, we would like to say that our present use of $G_2$ structures (3-forms in 7D) is different from what one can find in the literature on Kaluza-Klein compactifications of supergravity, see e.g. \cite{Duff:1986hr}. In this context, a 7D manifold with a $G_2$ structure is used for compactifying the 11D supergravity down to 4D. In contrast, here we propose to describe the gravity itself using 3-forms. In our approach a 3-form is not an object that exist in addition to the metric -- it is the only object that exist. The metric, and in particular the 4D metric, is defined by the 3-form via (\ref{metric}). 

\bigskip
The author was supported by ERC Starting Grant 277570-DIGT. He acknowledges the hospitality of the Isaac Newton Institute for Mathematical Sciences, Cambridge. These ideas were envisaged during the programme "Twistors, gravity and amplitudes" hosted by INI. The author also acknowledges support of the Albert Einstein Institute, Golm (Potsdam). This work has been completed during a research stay at AEI. Discussions with Yannick Herfay and Yuri Shtanov on the topics of relation between 7D and 4D are gratefully acknowledged.

\end{document}